
\input harvmac.tex

\def \P1{${\bf P}^1$}
\def \Z2{${\bf Z}_2$}

\noblackbox
\Title{\vbox{\hbox{HUTP-95/A049}
\hbox{\tt hep-th/9512078}}}
{Instantons on D-branes}
\bigskip\vskip2ex
\centerline{Cumrun Vafa}
\vskip2ex
\centerline{\it  Lyman Laboratory of Physics, Harvard
University}
\centerline{\it Cambridge, MA 02138, USA}
\vskip2ex
\vskip2ex

\vskip .3in

We consider type IIA compactification on $K3$.  We show
that the instanton subsectors of the worldvolume of
 $N$ 4-branes wrapped around $K3$ lead to a Hagedorn
density of BPS states in accord
with heterotic-type IIA duality in 6 dimensions.
  We also find evidence that the correct framework to understand
the results of Nakajima on the appearance of affine Kac-Moody algebras on
the cohomology of moduli space of
instantons on ALE spaces is in the context
of heterotic-type IIA string duality.

\Date{\it {Dec. 1995}}
%

One of the motivations behind the recent proposals
of string dualities has been the appearance of
field theory dualities, and in particular the Montonen-Olive
duality for N=4 YM theory.  The existence
of field theory dualities is a necessary condition to any generalization
to string theory.  There seems to be, however, another
indirect, but perhaps deeper connection
between the string duality and $N=4$ YM duality.
In \ref\vw{C. Vafa and E. Witten, Nucl. Phys. {\bf B431} (1994) 3}\
 a strong coupling test of
Montonen-Olive S-duality was done.  In particular $SU(2)$, $N=4$ YM
on $K3$ was considered and it was found that the partition function
of the theory is modular, in accord with S-duality.
  However, there was a surprising
aspect to this computation: In the course of this computation
the oscillator partition function of bosonic string theory
made an unexpected appearance!
  Technically this arose by relating
the moduli space of $SU(2)$ instantons on $K3$ to the moduli
space of some number of points on
$K3$.  A similar coincidence that was noticed
was that the partition function of $n$ unordered points on $T^4$
 similarly gives the oscillator partition function
for superstrings.  It was suggested in \ref\vgas{C. Vafa,
{\it Gas of D-Branes and Hagedorn Density of BPS States}, HUTP-95/A042,
hep-th/9511088}\
that these coincidences are not accidental and related to string
dualities: type II-type II duality (on $T^4$)
or type IIA (on $K3$)--heterotic (on $T^4$) duality.  In particular
it was shown there that a gas of $0$-branes on $T^4$ in the presence of one
$4$-brane wrapped around $T^4$
form bound states in accordance with predictions from string duality,
where the relevant moduli space that arises is that of a number
of points on $T^4$.
This generalized the result of \ref\senone{A. Sen,
{\it A Note on Marginally Stable Bound States in Type II String Theory},
MRI/PHY/23-95, hep-th/95102229}
\ref\sentwo{A. Sen,
{\it U-duality and Intersecting D-branes}, MRI/PHY/23-95,
hep-th/9511026}\
 who
showed (the $T$-dual of the statement)
that one 0-brane and one 4-brane form a bound state with the
expected degeneracy.   For heterotic--type II duality,
the suggestion of \vgas\ was realized in
\ref\bsvt{M. Bershadsky, V. Sadov and C. Vafa, {\it D-Branes
and Topological Field Theories} hep-th/9511222
}\ where it was shown that again
the relevant moduli problem that arises, whether it be
for a gas of 0-branes in the presence of a 4-brane, or
a 2-brane wrapped around a 2-cycle of $K3$, is that of a symmetric
product of a number of $K3$'s.

Even though these results are a strong confirmation both of the
string dualities
 and the D-branes as the relevant solitons \ref\pol{J. Polchinski,
{\it Dirichlet-Branes and Ramond-Ramond Charges}, NSF-ITP-95-122,
hep-th/9510017} (see also the related work \ref\witb{E. Witten,
{\it Bound States Of Strings And $p$-Branes},
 hep-th/9510135 }\vgas \senone \sentwo \ref\bsvo{M. Bershadsky, V. Sadov
and C.Vafa,
{\it D-Strings on D-Manifolds}, HUTP-95/A035, IASSNS-HEP-95-77,
hep-th/9510225}\bsvt\ref\wism{E. Witten, {\it Small Instantons in String
Theory}, hep-th/9511030}\ref\mli{M. Li, {\it
Boundary States of D-Branes and Dy-Strings},
hep-th/9510161}\ref\bach{
C.
Bachas, {\it  D-brane dynamics}, hep-th/9511043}\ref\kleb{C. G. Callan, and I.
R. Klebanov,
{\it D-Brane Boundary State Dynamics}, hep-th/9511173}\ref\bsu{T. Banks and L.
Susskind, {\it
Brane - Anti-Brane Forces}, hep-th/9511194}\ref\anst{A. Strominger,{\it Open
P-Branes}, hep-th/9512059}\ref\tow{P. Townsend, {\it D-branes from M-branes},
hep-th/9512062}\ref\rsc{R.R. Khuri and R.C. Myers, {\it  Rusty Scatter Branes},
hep-th/9512061}) two issues remained, one real and one aesthetic:
  The first one was what happens if we consider other
configurations, for example what if we have $N$ wrapped 4-branes
instead of one? Of course by T-duality on $K3$, one could
map the problem to the cases already considered, but
one would also like to be able to tackle this problem directly.
It is quite interesting that the flavor of the field theory
questions changes
so dramatically once one uses $T$-duality.  This suggests
that $T$-duality can be used to learn about non-trivial field theory
properties, such as strong/weak duality in N=4 YM.
Another issue mainly aesthetic, is that it seemed that
we had stumbled upon bosonic string partition function in
\vw\ quite accidentally.  It would have been
much more satisfactory if
not only the mathematical structure
but also the {\it physics} of the moduli problem encountered be the same
for testing of both the string duality and the Montonen-Olive duality.
We shall argue in this paper that these two issues nicely
complement one another and show that
by considering the configuration
of $N$ 4-branes wrapped around $K3$ the physics of the counting of
the Hagedorn density of BPS states
will {\it coincide} with the computation done in \vw\ and
so that computation can also be reinterpreted as a test of type II-heterotic
string duality!  That the physics may be identical
is not surprising when we recall that $N$ wrapped 4-branes around
$K3$ will give rise to $U(N)$ gauge theory on $K3 \times \bf R$ where
${\bf R}$ plays the role of time.  In fact the number of supersymmetries
is such that the
static configurations
of the 4-brane is identical to $N=4$ YM on $K3$.

Our starting point is the observation \witb\ that since
the dynamics of N D-branes is captured by
the maximally supersymmetric $U(N)$ gauge theory on the
worldvolume of the D-brane, the
normalizable
ground states of this gauge theory  can be reinterpreted
as bound states of $N$ D-branes.  We will need to generalize this
observation for our purposes:  What is the interpretation of excited
states of this supersymmetric $U(N)$ gauge theory on the worldvolume
of the D-brane?  Clearly they should be identified with the excitation
of the bound states. These excitations are generically unstable.  However,
let us consider a BPS state of the $U(N)$ gauge theory.
This in particular means that it is a reduced supersymmetry multiplet
of the $U(N)$ theory. Since we can think of the supersymmetry of the
$U(N)$ theory as coming from the `pull back' of the supersymmetries
in spacetime in the presence of D-brane bound state, we can interpret
these states as bound states which correspond to excitations of
D-brane bound states which are BPS saturated and thus stable
\foot{A good analogy to keep in mind is the case of 1-branes,
were the BPS states of the worldsheet theory, in the Green-Schwarz
formulation, correspond to excited states of the string which are killed
by half of the supersymmetries on the worldsheet (say the left-moving
supersymmetries). Clearly these correspond also to BPS states in the spacetime
\ref\harda{A. Dabholkar and J. Harvey, Phys. Rev. Lett. {\bf 63} (1989) 478}.
 In the above we are generalizing
this observation to Dirichlet p-branes.}.
Clearly the size of the spacetime
supersymmetry multiplet the bound state will reside in
will depend on how many supersymmetries the corresponding
BPS state preserves in the $U(N)$ worldvolume theory.

If we have $N$ 4-branes wrapped around the $K3$, we are
dealing with a 5 dimensional $U(N)$ gauge theory on
$K3 \times {\bf R}$ where ${\bf R}$ is the time direction.
Thus the instantons on $K3$, which preserve half the supersymmetries
of the theory, can be viewed as the BPS states of this theory.
Consider the instanton number $k$ sector of this theory, which
should be viewed as a particular Hilbert space subsector of this
5-dimensional theory. The situation at hand is very similar
to that considered in \ref\bjsv{M. Bershadsky, A. Johansen, V.
Sadov and C. Vafa,
Nucl. Phys. {\bf B448} (1995) 166} .  There
are two possibilities that may arise:  For generic
point on moduli of instantons the gauge symmetry is completely
broken and we can integrate out the gauge field also in the ${\bf R}$
direction.  If this were the full story we would obtain as the effective
theory on ${\bf R}$ the sigma model on instanton moduli space
and the ground states would be identified with the cohomology
elements of instanton moduli space.
However it may happen that for some singular loci of moduli
space of instantons we have (partially) enhanced gauge symmetry
which would imply that we cannot ignore the dynamics of the gauge theory
on ${\bf R}$, and in particular there will be a mixture of guage theory
coupled in a non-trivial way to a sigma model (on the rest of the instanton
moduli space).  This would be very complicated to deal with.
In fact whether one is in the `good case' or `bad case' should
be reflected by the nature of the singularities of the instanton moduli
space.  In particular the `bad case' should be accompanied with singular
instanton moduli space, for which one does not have a canonical recipe
for repairing the singularity and that the left over gauge dynamics
on ${\bf R}$ would be the physical information needed for the resolution
of the singularity.
  This situation is indeed similar to what was encountered
in \bjsv\ were in the `good case' there was a smooth moduli
space (the Hitchin space) and in the bad case there were singularities
with no
canonical resolution.
The nature of singularities of instanton moduli space is thus a strong hint
whether one will have to include extra dynamics in addition to the sigma
model on the instanton moduli space.

Let us denote the moduli space of $SU(N)$ instantons on $K3$
with instanton number $k$ by ${\cal M}^N_k$.
The dimension of ${\cal M}^N_k$ is
\eqn\dimf{{\rm dim}({\cal M}_k)=4\big[ N(k-N)+1\big]}
Let us recall that $N$ D-branes will have a $U(N)$ gauge symmetry
on their worldvolume
\ref\poldai{J. Dai, R. G. Leigh and J. Polchinski, Mod. Phys. lett,
{\bf A4}
(1989) 2073}\ref\hor{P. Horava, Phys. Lett. {\bf B231} (1989) 251},
 where the $U(1)\subset U(N)$ is identified
with the center of mass degree of freedom of the D-brane \witb .
Since for a generic metric on $K3$ there are no $U(1)$ instantons
we can ignore the $U(1)$ dynamics, and simply concentrate on the $SU(N)$
part\foot{Since $U(N)$ is not quite the direct product of $U(1)$ and $SU(N)$
in general this means that there are `t Hooft electric and magnetic flux
allowed on the manifold correlated with the $U(1)$ electric and magnetic
flux.  This in fact was a key point in the considerations of \witb .
Here, since the $U(1)$ field is trivial we do not have any `t Hooft
$Z_N$ fluxes turned on.}.

Before getting into the detail of the analysis we would like
to know what charges the BPS states would carry, in order
to compare with the expected degeneracy of states based on string duality.
In particular we would like to compute $P^2/2$  where $P^2=P_R^2-P_L^2$
on the heterotic side is identified with the length squared of the
momentum vector and $P_R,P_L$ are
 right and left
momenta of a vector in
the Narain lattice $\Gamma^{4,20}$.  The degeneracy of the BPS state
(apart from the supersymmetry multiplet degeneracy)
is expected to be $d({P^2\over 2}+1)$ where $d(n)$ denotes
the degeneracy of $n$-th oscillator level of left-movers
of bosonic strings, i.e.
\eqn\gan{\eta(q)^{-24}=q^{-1}\sum d(n) q^n}
where $\eta$ is the Dedekind eta function.  By identifying
the $\Gamma^{4,20}$ with the integral cohomology of $K3$,
we see that a $\Gamma^{1,1}$ part of it will correspond to
0- and 4-dimensional cohomologies.  This in particular means
that if we have $N$ 4-branes and $M$ 0-branes, we would identify
the $P^2/2$ with $NM$.
To begin with we have $N$ 4-branes and no 0-branes, so one may
think we are in the situation corresponding to $P^2=0$.  This
is not the case.
It has been pointed out in \bsvt\ that in general
there are quantum corrections which lead to anomalous charges.
In particular it was argued there that a 4-brane on $K3$ will acquire
in addition a $-1$ 0-brane charge.  This arose because on the worldvolume
there is a correction on the worldvolume of the 4-brane
 (converting from type IIB considered
in \bsvt\ to type IIA)
$${1\over 48} \int A \wedge p_1 (R) $$
where $A$ is the 1-form RR field coupling to the 0-branes
and $p_1(R)$ is the first Pontryagin class made of the Riemann tensor.
  For $K3$ this is
$p_1= -48$ and so we get a net effect on ${\bf R} $
of integration $-\int A$, i.e. we have $-1$ unit of 0-brane
charge.  For $N$ 4-branes that we are considering we would thus get,
in addition to the 4-brane charge,
$-N$ units of 0-brane charge.  Since we
 are considering instanton number
$k$ of $SU(N)$ we should ask whether or not there are any other
relevant corrections which would lead to a different counting of the
0-branes.  Indeed there is.  The argument outlined in \bsvt\
actually predicts in addition a term of the form
$$\int A \wedge c_2(F)$$
where $c_2(F)$ is the second Chern class of the $SU(N)$
gauge field living on the 4-brane worldvolume.  In fact, as was
pointed out to us by Douglas, this term is indeed generated
\ref\mdo{M. Douglas, {\it Branes within Branes}, hep-th/9512077} (see also
\ref\cet{C.G. Callan, C. Lovelace, C.R. Nappi and S.A. Yost,
Nucl. Phys. {\bf B308} (1988) 221}\mli ).
Noting that $c_2(F)=k$ is the instanton number of the $SU(N)$
gauge group, we learn that in the instanton number $k$ sector of the
$SU(N)$ gauge theory we have in addition $k$ units of the
0-brane charge. Thus the total effective 0-brane charge $M$, taking
into account the above corrections\foot{The relative sign
contributions has not been checked but, just as in \ref\vwo{C. Vafa and
E. Witten,
Nucl. Phys. {\bf B447} (1995) 261-270},
it should be so in order
to be consistent with string duality},
 is $M=k-N$. We thus have
\eqn\psq{{1\over 2}P^2=NM=N(k-N)}

Before getting into the analysis of the moduli space of instantons,
we can anticipate the `good' and `bad' cases based on what we have just
found.  One sign of the
 `bad' cases would be finding no mass gap. This will be
the case if $N$ and $k$ are not relatively prime.  To see this note that
if $N$ and $k$ are not relatively prime then $N$ and $M=k-N$
are also not relatively prime, which implies that the
BPS state we are about to find does not correspond to a primitive
charge in the  $\Gamma^{20,4}$ lattice.  Thus it has the same
energy as two subsystems.  Since these subsystems can be
separated to arbitrary distances we thus have no mass gap
and we would be up against finding a bound state at threshold.
This lack of mass gap should be reflected in the severe singularities
of moduli
space of $SU(N)$ instantons with instanton number $k$ not
relatively prime to $N$.  On the other hand if $k$ and $N$
are relatively prime, we should expect a mass gap and so the corresponding
moduli space of instantons has to have a discrete spectrum near the origin.
This suggests that in such cases the instanton moduli space should be
smooth enough to allow this.  So we have tentatively identified
the `good' and `bad' cases by checking whether $k$ and $N$
are relatively prime (good) or not (bad).
Below we will find
some indications which go in the direction of confirming this picture.

The moduli space of instantons on $K3$ is
 most intensively studied for $SU(2)$ (see \vw\ for the relevant
references).  From the above discussion we thus expect
that in this case for $k$ odd we get a good moduli space.
This is indeed the case and we find that
the cohomology of ${\cal M}^2_k$
for $k$ odd is the same as
the cohomology of $2k-3$ unordered points on $K3$, and moreover
for some choice of complex structure on $K3$ the moduli space
itself coincides with
$${\cal M}^2_k=(K3)^{2k-3}/S_{2k-3}$$
The Euler character of this space, as noted in \vw\ is $d(2k-3)$
(where $d(n)$ was defined in \gan ).  From string duality
we should have expected
$$d({1\over 2}P^2 +1)=d(2(k-2)+1)=d(2k-3)$$
which is thus a confirmation of the prediction of
heterotic-type II string duality.  For $k$ even,
the above discussion suggests that the moduli space
should be badly singular and the extra gauge
dynamics will have to be taken into account for its resolution.
That the space is singular is well known mathematically.  In fact
this was one of the difficulties in fully checking the prediction
of Montonen-Olive duality on $K3$ studied in \vw .  The $N=4$
duality in this case predicts a form for the Euler characteristic
of all the even cases as well, but the number is fractional \vw , again
indicating that the moduli space is singular.  A fractional
number cannot be counting the number of bound states.  In fact
as discussed above we expect
the string theory computation of bound states at threshold
 to differ from the computation of the
Euler characteristic of this space.

How about for other values of $N>2$?  Unfortunately this space
is not well studied mathematically.  But it is natural to conjecture
the following
(which was also conjectured in \vw\ based on $N=4$ duality):
The space ${\cal M}^N_k$ for $N$ and $k$ relatively prime is
`smooth' ( or more precisely has a canonical smooth resolution)
and its cohomology is the same\foot{Since by $T$-duality we can
convert a 4-brane to 0-brane, i.e. a point on $K3$, it seems
that this mathemetical conjecture may have a simple physical
interpretation.   In particular the proof in the case of $SU(2)$
involves relating instanton moduli to points on $K3$ which can be viewed
as the zeros of a canonical zero mode of a field transforming
in the fundamental of $U(2)$.  Since for a 0-brane and a 4-brane
a field in the fundamental naturally arises \bsvo\
 this mathematical
construction may have a natural physical interpretation.}
 as that of $N(k-N)+1$ unordered points
on $K3$.  First of all note that the dimension of the space of
$N(k-N)+1$ points on $K3$ agrees with the dimension of the moduli
space ${\cal M}^N_k$ given in \dimf .   Moreover, if this conjecture
is true it gives the correct prediction for the degeneracy of the
bound state, namely we expect to get $d(NM+1)=d(N(k-N)+1)$
as the degeneracy and this is exactly the cohomology of the space of
$N(k-N)+1$ unordered points on $K3$.
The strongest
evidence for this conjecture, apart from the fact that
it would be a natural generalization from the $SU(2)$ case,
 is that it is in fact true at least for all instanton numbers
$k=1$ mod $N$ \ref\kron{P. Kronheimer, private communication.}.

Let us note very briefly the situation with type II-type II
duality on $T^4$.  In this case the dimension of the moduli
of $U(N)$ instantons with instanton number $k$ is $4kN+4$ (taking into account
the $U(1)$ flat bundle on $T^4$).
 Moreover the effective 0-brane charge
is $k$ (as $p_1$ of $T^4$
is zero we do not get an additional
$-N$ that we got for $K3$).  To be in accord with string duality
(using the observation in \vw\ that the cohomology of $n$ unordered points
on $T^4$ agrees with the superstring oscillator degeneracy at level $n$)
it is natural to conjecture that if $k$ and $N$ are relatively prime the moduli
space of $SU(N)$ instantons is (or more precisely has the same cohomology as)
 the $kN$ fold symmetric product of $T^4$,
and together with the $U(1)$ flat moduli we would thus have the moduli
space $T^4 \times (T^4)^{kN}/S_{kN}$.  It would be interesting
to check this conjecture.

Let us close by making one remark in
 connection with the work of Nakajima \ref\nakajima{H. Nakajima, {\it
Homology of moduli
spaces of instantons on ALE Spaces. I}, J. Diff. Geom.
{\bf 40}(1990) 105; {\it Instantons on ALE spaces,
quiver varieties, and Kac-Moody algebras,} preprint;
{\it Gauge theory on resolutions of simple singularities
and affine Lie algebras}, preprint.}\
on instantons on ALE spaces.   There has already
been some discussion of this work in the physics literature \vw
\ref\moret{A. Losev, G. Moore, N. Nekrasov and S. Shatashvili,
{\it Four-Dimensional Avatars of Two-dimensional RCFT},
 Talks given at Strings 95 and
ICTP Conf. on S Duality and Mirror Symmetry, hep-th/9509151 }.  Nakajima shows
that if we consider $U(N)$ guage group on an ALE space, say
$A_{r-1}$ for concreteness, we get an action of $SU(r)$ Kac-Moody
algebra at level $N$ acting on the cohomology of $U(N)$
instanton moduli space.  In defining this action the
vanishing 2-cycles of the ALE play a crucial role.
It was suggested in \vw\ that a proper understanding of this
may involve at least a 5-dimensional theory, where the instanton
cohomologies would be viewed as BPS states and that string dualities
may play a role.  Here we would like to suggest that this is indeed
the case where we identify the missing 5-dimensional theory with
the worldvolume theory of the 4-brane.  In fact based on
string dualities \ref\ht{C. Hull and P. Townsend, Nucl. Phys. {\bf B360}
(1995) 109.}
\ref\witt{E. Witten, Nucl. Phys. {\bf B443} (1995) 85}
\bsvo\ and more directly using the D-branes
\bsvt\
one expects to have an enhanced gauge symmetry when a 2-cycle
wraps around a vanishing $S^2$.  The gauge particles
are identified with 2-branes wrapped around the 2-cycles.
For ALE corresponding to ADE we expect to get the corresponding
ADE gauge symmetry.  Now consider type IIA compactification
on a singular $K3$.  In this case the relevant singularity
can be modeled by an ALE space.  Now, by heterotic-type II
duality, we would land on a heterotic background near
an enhanced gauge symmetry point.  Moreover the instanton
cohomologies will arise from oscillator modes of heterotic strings.
  Thus the action
of the 2-branes on the instanton cohomology space can be
reinterpreted on the heterotic side as the action of the
affine Kac-Moody current algebra on the string oscillator states!
Note that this is in accord with the fact that the currents
on the heterotic side are left-movers and the BPS states
all come from the left-movers tensored with the ground
state of the right-movers.
The puzzle at first sight may be that we expect that the
heterotic side leads to a vertex operator algebra at level
1 and not $N$. There is a natural resolution of this puzzle:
As discussed above, $N$ 4-branes will {\it not} give
all the oscillator states, but all the oscillator levels separated
by $N$ units ($N_L=N\cdot M+1$ where $M$ varies).  Is there
any subalgebra of the Kac-Moody algebra which acts on this subspace?
Indeed there is:  Consider $J_n$ to be the current
of the affine Kac-Moody algebra.  Consider $\hat J_n=J_{nN}$.
Since
we have modified the moding by $N$ times the basic quantum
of oscillation this subalgebra can now act on the
cohomology of instanton moduli space of $SU(N)$ which
have oscillation gaps of $N$ units.
Moreover, it is an easy exercise to see that if the
original affine Lie algebra has level 1, as is predicted
by string duality,
the $\hat J_n$
form an affine Lie algebra, but now at level $N$.  We find this a strong hint
that the natural interpretation
of Nakajima's results is in the context of type II-heterotic
string duality.

We would like to thank M. Bershadsky, M. Douglas,
V. Sadov and S. Sethi
for valuable discussions.  The research
of C.~V.~is supported in part by NSF grant PHY-92-18167.

\listrefs

\end